\documentclass[12pt]{article}

\usepackage[english]{babel}
\usepackage{graphicx}   
\usepackage{ae,aecompl}

\title{Sliding rope paradox}

\author{Z.~K.~Silagadze \\
Budker Institute of Nuclear Physics and \\
Novosibirsk State University, 630 090, Novosibirsk, Russia }

\date{}

\begin{document}

\maketitle

\begin{abstract}
A simple mechanical problem is considered which we believe will help students 
to familiarize some concepts of mechanics of variable mass systems. Meanwhile
they can even learn some thrilling physics of bungee jumping.
\end{abstract}

\section{Introduction}
Sliding ropes and falling chains are often used in introductory mechanics
course as examples of variable mass systems \cite{1,2,3}. The problems at 
first seem deceptively simple but at closer inspection they aren't such.
An interesting history of falling chain problem \cite{4,5} reveals 
a considerable amount of both insight and confusion which often pervades 
such kind of problems. Conceptual difficulties of variable mass systems
even was considered \cite{6} as an argument against their inclusion at all
in any introductory mechanics course. However, variable mass systems play
such an important role in modern day technology that it is justified even
beginning students to be exposed to main principles of the mechanics of the
systems of variable composition. We agree with \cite{7} that the best way
to make the subject more palatable for beginning students is to introduce
it through a problem that allows complete enough treatment even without any 
reference to the variable mass systems and then, after the correct answers 
are already guessed, illustrate how the main principles of the variable 
mass systems can illuminate the problem.

In this note we reconsider an old problem \cite{8} which, in our 
opinion, is ideally suited for goals to introduce the concept of momentum 
flux and illustrate its use for systems of variable composition.
 
\section{Sliding rope paradox: what is the correct weight of the rope?}
A folded heavy rope of mass $m$ and length $2L$ is suspended on a little 
frictionless peg so that its right arm is a bit longer. The rope is released
and begins to slide down under the action of gravity. What is the weight of 
the sliding rope as felt by the peg?

\begin{figure}[htb]
\centerline{\includegraphics[height=40mm]{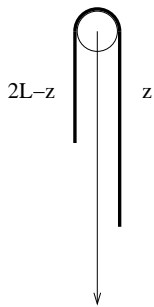}}
\caption{Folded heavy rope sliding on a peg.}
\label{Fig1}
\end{figure}

Let the length of the right arm of the rope is $z$ at some moment of time. 
When the $z$-component of the rope's momentum will be
\begin{equation}
p_z=\frac{m}{2L}z\dot{z}-\frac{m}{2L}(2L-z)\dot{z}=
\frac{m}{L}(z-L)\dot{z},
\label{eq1}
\end{equation}
where we have assumed that the $z$-axis is directed downward, see figure 
\ref{Fig1}.

\noindent We should have
\begin{equation}
\frac{d\,p_z}{d\,t}=mg-F,
\label{eq2}
\end{equation}
where $F$ equals just the weight of the sliding rope as felt by the supporting
peg. From (\ref{eq1}) and (\ref{eq2}) we get
\begin{equation}
F=mg-\frac{m}{L}\left [\dot{z}^2+(z-L)\,\ddot{z}\right ].
\label{eq3}
\end{equation}
Note that this last equation also follows from
$$m\ddot{z}_c=mg-F,$$
where
$$z_c=\frac{1}{m}\left [\frac{m}{2L}\,z\,\frac{z}{2}+\frac{m}{2L}\,(2L-z)\,
\frac{(2L-z)}{2}\right ]=\frac{1}{4L}\left [z^2+(2L-z)^2\right ]$$
is the center of mass coordinate of the rope. This is not surprising because 
for the rope, as a whole, the mass is not changed and therefore both forms
of Newton's equations,
\begin{equation}
\frac{d\vec{p}}{dt}=\vec{f}
\label{eq4}
\end{equation}
and
\begin{equation}
m\vec{a}=\vec{f},
\label{eq5}
\end{equation}
are equivalent.

The velocity of the rope, $\dot{z}$, can be found from energy conservation
$$m\frac{\dot{z}^2}{2}-mgz_c=\left . -mgz_c\right |_{z=L},$$
which gives
\begin{equation}
\dot{z}^2=\frac{g}{2L}\left [ z^2+(2L-z)^2-2L^2\right ]=\frac{g}{L}(z-L)^2.
\label{eq6}
\end{equation}
Differentiating (\ref{eq6}), we can find the acceleration $\ddot{z}$ of the 
rope
\begin{equation}
\ddot{z}=\frac{g}{L}\,(z-L).
\label{eq7}
\end{equation}
Substituting (\ref{eq6}) and (\ref{eq7}) into (\ref{eq3}), we get easily
\begin{equation}
F=mg\left [ 1-2\,\frac{(z-L)^2}{L^2}\right ].
\label{eq8}
\end{equation}
So far so good. However, we can think naively that 
\begin{equation}
F=2T,
\label{eq9}
\end{equation}
where $T$ is the tension of the rope near the peg. So (\ref{eq9}) opens 
a second easy way to find the sliding rope's weight. Namely, the equations 
of motion for the right and left parts of the rope are respectively
\begin{equation}
\frac{m}{2L}z\ddot{z}=\frac{m}{2L}zg-T
\label{eq10}
\end{equation}
and
\begin{equation}
-\frac{m}{2L}(2L-z)\ddot{z}=\frac{m}{2L}(2L-z)g-T.
\label{eq11}
\end{equation}
If (\ref{eq11}) is subtracted from (\ref{eq10}), we just get (\ref{eq7}) for
the rope's acceleration. The tension $T$ can be found from (\ref{eq10}), 
for example,
\begin{equation}
T=\frac{mg}{2}\,\frac{z}{L}\left(2-\frac{z}{L}\right )=\frac{mg}{2}
\left [1-\frac{(z-L)^2}{L^2}\right ].
\label{eq12}
\end{equation}
And here a big surprise is waiting for us: clearly (\ref{eq8}), (\ref{eq9})
and (\ref{eq12}) are not compatible! So what is the correct weight of the 
rope?
 
\section{Resolution of the paradox}
A first thought is that somehow 
(\ref{eq8}) is incorrect, because it predicts that the weight of the sliding
rope vanishes at
\begin{equation}
z=L+\frac{L}{\sqrt{2}},
\label{eq13}
\end{equation}
and beyond that even becomes negative! 

However, the truth is that both  (\ref{eq8}) and (\ref{eq12}) are correct, as
the logic leading to them was rock-solid. Only (\ref{eq9}) is at error. To 
understand why, let us trace the fate of the small piece of the rope that
disappears from the left arm just to join the right arm after an instant. The 
``instant'', however, has a finite duration $\Delta t=\pi R/\dot{z}$ if the
peg's radius is $R$ so that the length of the fold of the rope is $\pi R$. 
During $\Delta t$, a piece of the length $\pi R$ disappears from the left arm 
of the rope and an equivalent piece of the same length joins the right arm. 
However, the piece that disappeared from the left arm had upward velocity 
$-\dot{z}$, while the  velocity $\dot{z}$ of the part that joins the right 
arm points downward. Therefore, a small mass $\Delta m=\frac{m}{2L}\pi R$, 
which was transferred from one sub-rope to the other, experiences an effective 
downward acceleration
$$a_z=\frac{\dot{z}-(-\dot{z})}{\Delta t}=\frac{2\dot{z}}{\pi R}.$$
When the radius $R$ of the peg goes to zero, $a_z$ increases without bound.
However, $\Delta m$ goes to zero as well, so that $\Delta m \,a_z$ remains 
finite:
$$\Delta m \,a_z=\frac{m}{L}\,\dot{z}^2.$$
Therefore, from $\Delta m \,a_z=2T-F$ we get
\begin{equation}
F=2T-\frac{m}{L}\,\dot{z}^2=2T-mg\,\frac{(z-L)^2}{L^2},
\label{eq14}
\end{equation}
which is just the relation between (\ref{eq8}) and (\ref{eq12}).

To treat the problem more systematically, this is a good point to introduce  
variable mass systems and indicate that the root of our confusion, which led
to (\ref{eq9}), can be traced to the fact that for variable mass systems 
neither (\ref{eq4}) nor (\ref{eq5}) are, in general, valid equations of 
motion. Instead, the right starting point for variable mass systems is
\cite{6}
\begin{equation}
\frac{d\vec{p}}{dt}=\vec{f}+\vec{\Pi},
\label{eq15}
\end{equation}
where $\vec{\Pi}$ is the momentum flux into the system - how much momentum is 
brought into the system by new parts in unite time. The logic of (\ref{eq15})
is simple: the momentum of the system of variable composition changes not
only because of the action of the resultant external force $\vec{f}$, but 
also because new parts of the system can bring some momentum into the system.

Left and right arms of the rope can be considered as variable mass systems.
In unite time, new parts bring the momentum
\begin{equation}
\Pi=\frac{m}{2L}\dot{z}^2
\label{eq16}
\end{equation}
into the right arm. Parts that disappear from the left arm have negative 
momentum. Therefore the $z$-component of the momentum flux into the left arm
is also positive: as a result of disappearance of those parts from the left
arm  $z$-component of its momentum increases (becomes less negative). Then
(\ref{eq15}) indicates that the equations of motion for the right and left 
parts of the rope are respectively
\begin{equation}
\frac{d}{dt}\left (\frac{m}{2L}z\dot{z}\right )=\frac{m}{2L}zg-T+\Pi
\label{eq17}
\end{equation}  
and
\begin{equation}
\frac{d}{dt}\left (-\frac{m}{2L}(2L-z)\dot{z}\right )=\frac{m}{2L}(2L-z)g
-T+\Pi.
\label{eq18}
\end{equation} 
Because of (\ref{eq16}), the equations (\ref{eq17}) and (\ref{eq18}) are
equivalent to our old equations (\ref{eq10}) and (\ref{eq11}). However, if we
add (\ref{eq17}) and (\ref{eq18}), we get
$$\frac{d}{dt}\left (\frac{m}{L} (z-L) \dot{z}\right )=mg-2T+2\Pi,$$
and comparing with (\ref{eq2}), we see that
$$F=2T-2\Pi,$$
which is just the equation (\ref{eq14}).

\begin{figure}[htb]
\centerline{\includegraphics[height=30mm]{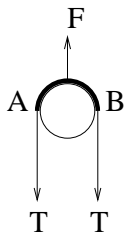}}
\caption{Fold of the rope and forces acting on it.}
\label{Fig2}
\end{figure}
At last, let us consider the equation of motion for the fold $AB$ of the rope
(figure \ref{Fig2}). Its center of mass remains at rest. Therefore 
$\frac{dp_z}{dt}=0$ for the fold. The mass $\Delta m=\frac{m}{2L}\pi R$ of 
the fold remains also unchanged. However we have non-zero momentum flux  
into the fold. The pieces of the rope which join the fold in unite time bring 
the negative momentum $-\Pi$ with them, while the parts that disappear from 
the fold in unite time have in total the positive momentum $\Pi$. Therefore 
the total momentum flux into the fold is negative and equals to $-2\Pi$. 
Hence the equation of motion of the fold is
$$0=\frac{dp_z}{dt}=\Delta m\, g+2T+\Delta T-F-2\Pi,$$
which gives again the equation (\ref{eq14}) in the limit $R\to 0$ (in fact,
rope tensions differ slightly at the left and right ends of the bend - hence
the term $\Delta T$ in the above equation. However $\Delta T\to 0$ when
$R\to 0$. For finite $R$, the calculation of $\Delta T$ is described in 
\cite{8}).

Let us emphasize that (\ref{eq15}) is the correct equation of motion for 
systems of {\it variable composition}, the variable mass systems being just 
a particular case of such systems. Our last example above, the equation of 
motion of the rope's fold, shows clearly that the momentum flux may matter 
even in the case of constant mass systems, if their composition varies over 
time.

\section{What about negative weight?}
As was already mentioned, (\ref{eq8}) indicates that the weight of the 
sliding rope vanishes when the coordinate $z$ of the rope's right end becomes
$$z=L+\frac{L}{\sqrt{2}}.$$
At that instant the velocity of the rope that follows from (\ref{eq6}) is
\begin{equation}
v=\sqrt{\frac{gL}{2}}.
\label{eq19}
\end{equation}
What happens next actually depends on the arrangement how the rope is fixed on
the peg \cite{9}. If it goes through a frictionless channel in the peg, like
a small duct, the negative weight which follows from (\ref{eq8}) will become
a reality. However, if the rope is not fixed in this manner on the peg, so 
that the arrangement is more like a pulley than a duct, the rope will start
to whiplash after its velocity reaches (\ref{eq19}).

Assuming an ideal perfectly flexible rope, the motion will remain effectively
one-dimensional nevertheless and can be analytically described \cite{8}.
Moreover, this can be performed without using variable mass systems, after the
rope's tension at the bend and the speeds of the ends of the rope are related
by an educative guess \cite{8}. The concept of momentum flux, however, allows 
to explain this relation for the tension more naturally. 

\begin{figure}[htb]
\centerline{\includegraphics[height=40mm]{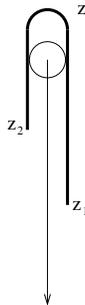}}
\caption{The rope after it leaves the peg.}
\label{Fig3}
\end{figure}

Figure \ref{Fig3} shows the rope after the bend in the rope begins to move 
away from the peg. The coordinates of the right and left ends of the rope
are $z_1$ and  $z_2$ respectively, while the coordinate of the 
(infinitesimally small) bend itself is $z$ (note that $z<0$). As the rope is
inextensible, we should have
\begin{equation}
z_1+z_2-2z=2L
\label{eq20}
\end{equation}
and, therefore, the bend moves with velocity
\begin{equation}
\dot{z}=\frac{1}{2}(\dot{z}_1+\dot{z}_2).
\label{eq21}
\end{equation} 
In the non-inertial frame $S$, where the bend is at rest, the right arm of 
the rope moves downward with velocity
\begin{equation}
v_1=\dot{z}_1-\dot{z}=\frac{1}{2}(\dot{z}_1-\dot{z}_2),
\label{eq22}
\end{equation}
while the left arm moves upward with velocity
$$v_2=\dot{z}_2-\dot{z}=\frac{1}{2}(\dot{z}_2-\dot{z}_1)=-v_1.$$
Therefore the situation in this system is analogous to the one considered in 
the previous section and hence we have momentum flux $-2\Pi$ into the fold, 
where now
\begin{equation}
\Pi=\frac{m}{2L}\,v_1^2=\frac{m}{2L}\,\frac{1}{4}\,(\dot{z}_1-\dot{z}_2)^2.
\label{eq23}
\end{equation}
In the limit $R\to 0$, the equation of motion of the fold becomes (note that
the inertial force $-\Delta m\, \ddot{z}$ also goes to zero in this limit,
along with $\frac{dp_z}{dt}$, $\Delta m\,g$ and $\Delta T$)
$$0=2T-2\Pi,$$
which indicates that
\begin{equation}
T=\Pi=\frac{m}{2L}\,\frac{1}{4}\,(\dot{z}_1-\dot{z}_2)^2.
\label{eq24}
\end{equation}
In the system $S$, equations of motion for the right and left arms of the 
rope take the form
\begin{equation}
\frac{d}{dt}\left [ \frac{m}{2L}(z_1-z)(\dot{z}_1-\dot{z})\right ]=
\frac{m}{2L}(z_1-z)g-T+\Pi-\frac{m}{2L}(z_1-z)\ddot{z}
\label{eq25}
\end{equation}
and
\begin{equation}
\frac{d}{dt}\left [ \frac{m}{2L}(z_2-z)(\dot{z}_2-\dot{z})\right ]=
\frac{m}{2L}(z_2-z)g-T+\Pi-\frac{m}{2L}(z_2-z)\ddot{z}.
\label{eq26}
\end{equation}
The last terms in (\ref{eq25}) and (\ref{eq26}) represent inertial forces.
Because of  (\ref{eq23}), this equations are equivalent to simpler equations
\begin{equation}
\frac{m}{2L}(z_1-z)\ddot{z}_1=\frac{m}{2L}(z_1-z)g-T
\label{eq27}
\end{equation}
and
\begin{equation}
\frac{m}{2L}(z_2-z)\ddot{z}_2=\frac{m}{2L}(z_2-z)g-T.
\label{eq28}
\end{equation}
After this point, the solution is essentially given in \cite{8} and goes
as follows. Let us introduce the relative coordinate
\begin{equation}
z_m=z_1-z_2,
\label{eq29}
\end{equation}
and note that the center-of-mass position is given by
\begin{equation}
z_c=\frac{1}{m}\left [\frac{m}{2L}(z_1-z)\frac{z_1+z}{2}+
\frac{m}{2L}(z_2-z)\frac{z_2+z}{2}\right ]=\frac{z_1+z_2}{2}+
\frac{z_m^2}{8L}-\frac{L}{2},
\label{eq30}
\end{equation} 
where we have used $z=(z_1+z_2)/2-L$ from (\ref{eq20}). Equations 
(\ref{eq29}) and (\ref{eq30}) enable to express $z_1$ and $z_2$ through $z_m$
and  $z_c$:
\begin{equation} 
z_1=z_c+\frac{L}{2}+\frac{z_m}{2}-\frac{z_m^2}{8L}, \;\;\;
z_2=z_c+\frac{L}{2}-\frac{z_m}{2}-\frac{z_m^2}{8L}.
\label{eq31}
\end{equation}  
After the rope leaves the peg, the center of mass experiences a free fall
with acceleration $g$. At the moment the whiplash begins, 
$$z_1=L+\frac{L}{\sqrt{2}},\;\;\; z_2=2L-z_1=L-\frac{L}{\sqrt{2}},$$
and we find from (\ref{eq30})
$$\left . z_c\right |_{t=0}=\frac{3}{4}L.$$
While (\ref{eq6}) indicates that
$$\left . \left . \dot{z}_1\right |_{t=0}=-\dot{z}_2\right |_{t=0}=
\sqrt{\frac{gL}{2}},$$
and hence
$$\left . \dot{z}_c\right |_{t=0}=\left [ \frac{\dot{z}_1+\dot{z}_2}{2}+
\frac{z_1-z_2}{4L}(\dot{z}_1-\dot{z}_2)\right ]\left . \right |_{t=0}=
\frac{1}{2}\sqrt{gL}.$$
Therefore, the center-of-mass motion is given by
\begin{equation}
z_c=\frac{3}{4}L+\frac{1}{2}\sqrt{gL}\, t+\frac{1}{2}gt^2,
\label{eq32}
\end{equation}
where the time $t$ is measured from the instant the rope leaves the peg.

For the relative coordinate $z_m$, we get from (\ref{eq27}) and (\ref{eq28})
$$\ddot{z}_m=-\frac{2L}{m}T\left [\frac{1}{z_1-z}-\frac{1}{z_2-z}\right ]=
\frac{1}{4}\,\frac{\dot{z}_m^2z_m}{(z_1-z)(z_2-z)}.$$
But
$$ (z_1-z)(z_2-z)=\left [\frac{1}{2}(z_1-z_2)+L\right ]
\left [-\frac{1}{2}(z_1-z_2)+L\right ]=L^2-\frac{1}{4}\,z_m^2.$$
Therefore, finally
\begin{equation}
\ddot{z}_m=\frac{z_m\dot{z}_m^2}{4L^2-z_m^2}.
\label{eq33}
\end{equation}
This equation for $z_m$ seems formidable but, in fact, it is not as terrible 
as it looks. From (\ref{eq33}) we have 
$$\frac{d}{dt}\left [ (4L^2-z_m^2)\dot{z}_m^2\right ]=0.$$
Therefore,
\begin{equation}
\sqrt{4L^2-z_m^2}\,\dot{z}_m=\left . \sqrt{4L^2-z_m^2}\,\dot{z}_m \right 
|_{t=0}=2L\sqrt{gL}.
\label{eq34}
\end{equation}
If now we introduce a new variable $\phi$ through
\begin{equation}
z_m=2L\sin{\phi},
\label{eq35}
\end{equation}
equation (\ref{eq34}) will take the form
\begin{equation}
(1+\cos{2\phi})\,\dot{\phi}=\sqrt{\frac{g}{L}}.
\label{eq36}
\end{equation}
This equation can be easily integrated (note that $\left . \phi \right 
|_{t=0}=\pi/4$, because $\left . z_m \right |_{t=0}=\sqrt{2}L$) with the
result
\begin{equation}
2\phi+\sin{2\phi}=\frac{\pi}{2}+1+2\sqrt{\frac{g}{L}}\, t.
\label{eq37}
\end{equation} 
This equation implicitly determines the function $\phi(t)$ and hence, 
through (\ref{eq35}), the function $z_m(t)$. Relations (\ref{eq31}), 
together with (\ref{eq32}), then completely determine the motion of the rope
up to the time
$$t=\sqrt{\frac{L}{g}}\,\frac{1}{2}\,\left (\frac{\pi}{2}-1\right ),$$
when the rope becomes straight and the fold disappears (note that at this 
moment $z_m=2L$ and, therefore, $\phi=\pi/2$).

\section{Connections with falling chain}
Our considerations of sliding rope above can shed light also to another 
interesting problem, the problem of falling chain \cite{4,5}. One end of 
a folded flexible chain is fixed on a rigid support. The another end is 
released in the manner of a bungee fall. If initially the ends were placed 
close together, the chain motion is effectively one-dimensional and 
conceptually much like to the previous case of sliding rope.  

Whether the falling part of the chain is in a free fall and moves with the 
acceleration $g$, surprisingly was a subject of controversy for a long time 
\cite{4,5}. It seems the overall consensus was in favor of the free fall. 
Sommerfeld gives a convincing argument \cite{2} (albeit for an another 
falling chain problem, the steady link-by-link fall of a stationary chain 
from a resting heap), that the energy is not conserved in such problems due 
to inelastic collisions when new links are brought in motion. Therefore, one 
can expect that the falling arm is freely falling with acceleration $g$ and 
is gradually brought to rest link by link by inelastic impacts at the chain 
fold. However, the last word in physics is an experiment, and the experiments 
\cite{10,11,12}, including the ones modeling bungee jumping \cite{13,14}, 
indicate that the tip of the falling chain moves with acceleration grater 
than $g$. What a surprise, this means that there is a downward pull at the 
fold even when the bungee rope is still slack! ``This result is contrary to 
the usual experience with free falling objects and therefore hard to believe 
for many a person, even an experienced physicist'' \cite{14}. However, the 
concept of momentum flux allows to explain the presence of this tension at 
the fold as naturally as it was in the case of sliding rope.

\begin{figure}[htb]
\centerline{\includegraphics[height=40mm]{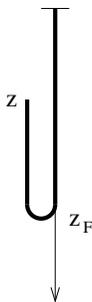}}
\caption{The falling bungee rope.}
\label{Fig4}
\end{figure}

Let the coordinates of the falling tip of the rope and of the fold are $z$ and 
$z_F$ respectively (figure \ref{Fig4}).

\noindent Then $z+2(z_F-z)=2L$, which gives
\begin{equation}
z_F=L+\frac{z}{2},\;\;\; \dot{z}_F=\frac{\dot{z}}{2}.
\label{eq38}
\end{equation}
In the non-inertial frame $S$, where the fold is at rest, the left arm of the 
rope has the downward velocity
$$v_1=\dot{z}-\dot{z}_F=\frac{\dot{z}}{2},$$
while the right arm moves upward and its velocity is 
$$v_2=0-\dot{z}_F=-v_1.$$
Therefore, in the frame $S$, we have a momentum flux $2\Pi$ in the fold with
\begin{equation}
\Pi=\frac{m}{2L}\,v_1^2=\frac{m}{2L}\,\frac{1}{4}\,\dot{z}^2.
\label{eq39}
\end{equation}
In its rest frame, equation of motion of the fold looks like 
\begin{equation}
0=\Delta m\,g-2T-\Delta T+2\Pi-\Delta m\,\ddot{z_F},
\label{eq40}
\end{equation} 
and in the limit of the vanishing radius of the fold this equation reduces to
\begin{equation}
T=\Pi=\frac{m}{2L}\,\frac{1}{4}\,\dot{z}^2.
\label{eq41}
\end{equation} 
We could stop here, as (\ref{eq41}) is just the mysterious downward pull in 
the bungee fall \cite{1}. However, for completeness reasons, we proceed and 
re-derive some further results for falling chain.

Equations of motion of the left and right subchains are (in the non-inertial
system $S$ and in the limit of vanishing radius of the fold)
\begin{eqnarray} & &
\frac{d}{dt}\left [ \frac{m}{2L}(2L-z_F)(\dot{z}-\dot{z}_F)\right]=
\frac{m}{2L}(2L-z_F)g+T-\Pi-\frac{m}{2L}(2L-z_F)\ddot{z}_F, \nonumber \\ & &
\frac{d}{dt}\left [ \frac{m}{2L}z_F(0-\dot{z}_F)\right]=
\frac{m}{2L}z_Fg+T-F-\Pi-\frac{m}{2L}z_F\ddot{z}_F,
\label{eq42}
\end{eqnarray}
where $F$ stands for the weight of the falling chain. Equations (\ref{eq42})
simplify to
\begin{equation}
\frac{m}{2L}\left (L-\frac{z}{2}\right )\ddot{z}=\frac{m}{2L}
\left (L-\frac{z}{2}\right )g+T
\label{eq43}
\end{equation}
and
\begin{equation}
F=\frac{m}{2L}\left (L+\frac{z}{2}\right )g+T.
\label{eq44}
\end{equation}
Because of (\ref{eq41}), equation  (\ref{eq43}) takes the form
\begin{equation}
\left (L-\frac{z}{2}\right )\ddot{z}=\left (L-\frac{z}{2}\right )g+
\frac{1}{4}\dot{z}^2.
\label{eq45}
\end{equation}
Again, this non-linear differential equation looks hard to solve, but note 
the following relation which follows from it:
$$\frac{d}{dt}\left [\left (L-\frac{z}{2}\right )\dot{z}^2\right ]=
2\dot{z}\left [\left (L-\frac{z}{2}\right )\ddot{z}-\frac{1}{4}\dot{z}^2
\right ]=$$ $$2\dot{z}g\left (L-\frac{z}{2}\right )=\frac{d}{dt}\left [
2gz\left (L-\frac{z}{4}\right )\right ].$$
Therefore, (\ref{eq45}) has the following first integral (note the initial 
condition $\dot{z}=0$, when $z=0$)
\begin{equation}
\left (L-\frac{z}{2}\right )\dot{z}^2=2gz\left (L-\frac{z}{4}\right ).
\label{eq46}
\end{equation}
In fact, (\ref{eq46}) represents energy conservation \cite{1}. Therefore,
the velocity of the falling subchain is given by
\begin{equation}
\dot{z}=\sqrt{gz\,\frac{4L-z}{2L-z}}.
\label{eq47}
\end{equation}
Then from (\ref{eq45}) we get its greater than $g$ acceleration
\begin{equation}
\ddot{z}=g\left [1+\frac{z(4L-z)}{2(2L-z)^2}\right ],
\label{eq48}
\end{equation}
while (\ref{eq44}) gives the weight of the falling chain \cite{4}
\begin{equation}
F=mg\,\frac{8L^2-3z^2+4Lz}{8L(2L-z)}.
\label{eq49}
\end{equation}
These results for the acceleration and weight can be generalized to include
the mass of the subject attached to the falling tip of the chain (or rope)
\cite{13,14}. The finite mass of the bungee jumper regularizes (\ref{eq47}),
(\ref{eq48}) and (\ref{eq49}), removing unphysical infinities in the velocity,
acceleration and weight when $z\to 2L$. The reader can easily modify our 
formulas above and reproduce these more general results.

\section{Concluding remarks}
We think, sliding rope problem is a good starting point to introduce the 
basics of variable mass mechanics to beginning students. The problem looks
simple but at closer inspection reveals some non-trivial features. Therefore, 
to disentangle it, we hope, will be both interesting and informative 
enterprise for students. The machinery of variable mass mechanics, learned 
through this problem, can be applied to other similar situations. For example, 
students can analyze falling chains and ropes and discover a surprising fact 
that bungee jumper falls with acceleration greater than $g$, even though 
bungee rope is still slack in the first phase of bungee jumping.

That a slack bungee rope can cause a downward pull may be surprising for 
students and even for experienced physicist \cite{14}. However, to prove the
reality of this pull, one should have not necessarily to go through a 
thrilling experience of real bungee jumping. Einstein's equivalence principle 
suffices to design a simple, elegant and convincing demonstration 
\cite{1,10,15}.

Place a folded chain, or rope, on a smooth table and quickly pull one end of 
the rope. You will find that the other end begins to move toward the bend,
opposite to the direction of the pull. Therefore, there must exist a tension 
in the rope at the fold that drags the free end toward the bend. However, in
the non-inertial frame, where the yanked end is at rest, we have an effective
gravity and the situation is just that is involved in the falling chain
problem.

This demonstration also indicates that similar physics is involved in
whip\-cracking. However, in latter case the size of the fold is not small and
the motion is no longer effectively one-dimensional. Therefore, more
advanced methods are needed to investigate this fascinating phenomenon
\cite{16,17,18}, possibly discovered by dinosaurs millions of years ago
\cite{19}.

It is surprising that without experiments our intuition can go astray even
in relatively simple situations, like the falling chain problem. First 
experiments \cite{20}, convincingly demonstrating a counter-intuitive result 
that the tip of the falling chain falls with acceleration greater than $g$, 
have not had attracted any significant attention and this remarkable
fact was largely ignored and unknown until new experiments \cite{10}
have confirmed it.  

Modern physics operates with space, time and matter at the Planck scale that
cannot be directly probed by experiment in the foreseeable future. Therefore,
one faces the question ``Can there be physics without experiments?'' 
\cite{21}. The history of falling chain suggest that the answer is firmly 
``No, there can not be any physics without experiments''. Without experiments
we are always in a danger of being deceived by our metaphysical 
misconceptions. For example, although variable mass systems were under 
investigation for more than two hundred years, unexpected and unexplainable
anomalous behavior was observed, revealing flaws in current understanding of 
the dynamics of variable mass systems, when a new class of solid rocket motors 
was used in early 1980's to power upper stages of several space missions
\cite{22}.

As the last remark, note that the falling chain history can serve as an 
illustration of subtle interplay between theoretical and experimental methods 
in physics. There were two theoretical paradigms in falling chain problem:
energy conserving and the one that assumes maximal kinetic-energy loss due to 
completely inelastic collisions. It may seem at first sight that the first
paradigm was vindicated by experiment and the dispute is over. However, in 
any theoretical construction one deals with some idealized model of reality, 
real ropes and chains being somewhere in between of those two extreme 
idealizations. 

\begin{figure}[htb]
\centerline{\includegraphics[height=40mm]{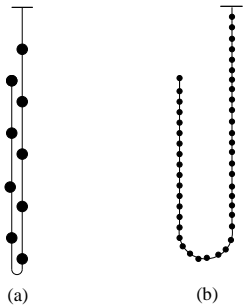}}
\caption{Two limiting cases of flexible chain \cite{1}.}
\label{Fig5}
\end{figure}

An ideal, completely flexible, infinitesimally thin and unstretchable chain
can be modeled as a series of point masses connected by massless strings.
The behavior of such ideal folded chain, however, depends on the relative
size of two specific length scales implicit in the model \cite{1}. If the 
spacing between the point masses is large compared with the horizontal size
of the fold (figure \ref{Fig5}(a)), each of the masses stops abruptly in an
inelastic collision when it reaches the singularity (kink) in the rope. The 
tension is not continuous along the kink, so that $\Delta T$ in (\ref{eq40})
does not vanish for infinitesimal bend and our previous analysis  breaks at 
this point.

On the contrary, if the spacing between the point masses is small compared 
with the horizontal span of the bend (figure \ref{Fig5}(b)), each mass 
comes to rest gradually and the system is energy conserving.

Experiments \cite{10,11,12,13,14} indicate that the real ropes and chains 
involved in these experiments were more close to the energy conserving variant
of figure \ref{Fig5}(b). However, they say nothing about a possibility of 
experimental realization of another variant of figure \ref{Fig5}(a). Moreover,
this variant is more interesting, as an example of singular one-dimensional 
system with a kink. Therefore, it will be very illuminating to demonstrate
that this type of the extreme limiting case of flexible chain could be also 
realized experimentally.

\section*{Acknowledgments}
The author's attention to the problem of falling rope's weight was drawn by
D.~A.~Medvedev.

\end{document}